\documentstyle[pra,aps,psfig]{revtex}
\newcommand{\beq}{\begin{equation}}
\newcommand{\eeq}{\end{equation}}
\newcommand{\beqa}{\begin{eqnarray}}
\newcommand{\eeqa}{\end{eqnarray}}
\newcommand{\ba}{\begin{array}}
\newcommand{\ea}{\end{array}}

\begin{document}

\draft

%%%%%%%%%%%%%%%%%%
\twocolumn[\hsize\textwidth\columnwidth\hsize
\csname@twocolumnfalse\endcsname
%%%%%%%%%%%%%%%%%%

\widetext 

\title{Dynamics of a BEC bright soliton in an expulsive potential} 
\author{Luca Salasnich} 
\address{$^{1}$Istituto Nazionale per la Fisica della Materia, 
Unit\`a di Milano, \\
Dipartimento di Fisica, Universit\`a di Milano, \\
Via Celoria 16, 20133 Milano, Italy}

\maketitle

\begin{abstract} 
We theoretically investigate the dynamics of a 
matter-wave soliton created in a harmonic potential, which is 
attractive in the transverse direction but expulsive in the 
longitudinal direction. This Bose-Einstein-condensate (BEC) 
bright soliton made of $^7$Li atoms has been observed 
in a recent experiment (Science {\bf 296}, 1290 (2002)). 
We show that the non-polynomial Schr\"odinger equation, an 
effective one-dimensional equation we derived from 
the three-dimensional Gross-Pitaevskii equation, is able 
to reproduce the main experimental features of this  
BEC soliton in an expulsive potential. 
\end{abstract}


%%%%%%%%%%%%%%%%%%
]
%%%%%%%%%%%%%%%%%%

\narrowtext 

\newpage

\section{Introduction} 

Bright and dark solitons are localized waves that travel over 
large distances without spreading. Bright solitons are local maxima 
in the field while dark solitons are local minima. 
Solitons are ubiquitous: they appear in systems as diverse as 
oceans, shallow waters in narrow channels, 
electric circuits and optical fibers \cite{1}. 
Recently solitons have been experimentally produced 
in Bose-Einstein condensates (BECs) of alkali-metal atoms 
\cite{2,3,4,5}. 
Dark solitons have been 
obtained with repulsive $^{87}$Rb atoms \cite{2}, while bright solitons 
have been observed with attractive $^7$Li atoms \cite{3,4} 
and also in a optical lattice with $^{87}$Rb 
atoms (gap bright solitons) \cite{5}. 
\par 
BEC solitons have been theoretically investigated in various papers but 
only in few of them there is a detailed comparison 
with experiments \cite{6,7,8}. For instance, 
by using the Gross-Pitaevskii equation \cite{9}, in Ref. \cite {8}
we have successfully simulated the formation and dynamics of the train 
of bright solitons observed in the experiment of 
the Rice University \cite{3}. 
In this paper we consider instead the experiment of the Ecole Normale 
Sup\'erieure \cite{4}, where Khaykovich {\it et al.} have produced  
and detected a single BEC bright soliton in a expulsive 
potential. The stable configurations 
of this BEC bright soliton have been analyzed by Carr 
and Castin \cite{7} by means of an analytical variational method. 
Here we complete their theoretical investigation by 
studying numerically the full time-evolution of the bright soliton 
by using the non-polynomial Schr\"odinger equation (NPSE) \cite{10}, 
an effective one-dimensional (1D) equation we have recently derived 
from the 3D Gross-Pitaevskii equation \cite{9}. 
We find that NPSE reproduces remarkably well the experimental data, 
in particular the center of mass dynamics of the BEC cloud 
and also the time dependence of its longitudinal width. 

\section{Expulsive potential and NPSE} 

In the experiment of Khaykovich {\it et al.} \cite{4} the Bose-Condensed 
$^7$Li atoms are confined in a trap that can be modelled 
by a harmonic potential 
\beq 
U({\bf r}) = {1\over 2}m \left[ 
\omega_{\bot}^2 (x^2+y^2) + \omega_z^2 z^2 \right] \; ,  
\eeq 
where $m$ is the atomic mass, 
$\omega_{\bot} = 2\pi \times 710$ Hz is the transverse 
frequency and $\omega_z = 2\pi i \times 78$ Hz is the {\it imaginary}  
longitudinal frequency. This imaginary frequency is due to an 
offset magnetic field that produces the slightly expulsive 
harmonic potential $-(m/2)|\omega_z|^2 z^2$ 
in the longitudinal direction. 
The main effect of this expulsive term is that the center of mass 
of the BEC accelerates along the longitudinal direction \cite{4}. 
\par 
The dynamics of a BEC at zero temperature can be described by the 
time-dependent 3D Gross-Pitaevskii equation (3D GPE), given by 
\beq 
\left[ i\hbar {\partial \over \partial t} +{\hbar^2\over 2 m} 
\nabla^2 - U  - g N |\psi|^2 \right] \psi = 0  \; , 
\eeq 
where $\psi({\bf r},t)$ is the macroscopic wave function 
(order parameter) of the Bose-Einstein condensate \cite{9}, 
$g=4\pi\hbar^2 a_s/m$ is the interatomic strength, $a_s$ is the s-wave 
scattering length, and $N$ is the number of condensed atoms. 
\par 
In the case of strong cylindric radial confinement, 
i.e. when the BEC travsverse energy $E_{\bot}$ 
of the BEC is equal to the transverse 
harmonic energy $\hbar \omega_{\bot}$, the 3D GPE reduces to a 1D GPE. 
This result can be easily obtained with a variational approach 
by using the following trial wave function $\psi(x,y,z,t) = 
f(z,t) \exp{[-(x^2+y^2)/(2\eta^2)]}/\sqrt{\pi \eta^2}$, 
where $f(z,t)$ is the longitudinal wave function normalized to one and 
$\eta$ is the transverse width. Actually one finds 
$$
\left[ i\hbar {\partial \over \partial t} + {\hbar^2\over 2m} 
{\partial^2\over \partial z^2} - {\hbar \omega_{\bot}\over 2}
\left( {a_{\bot}^2\over \eta^2 } + {\eta^2 \over a_{\bot}^2} 
\right) - {m\over 2}{\omega_z^2} z^2 
\right. 
$$
\beq 
\left. 
- {gN \over (2 \pi) \eta^2} |f|^2 
\right] f = 0 \; ,    
\eeq  
where $E_{\bot} = (\hbar \omega_{\bot}/2)
(a_{\bot}^2/\eta^2  + \eta^2/a_{\bot}^2)$ 
is the transverse energy 
of the Bose condensate. In the 1D GPE one has $\eta = a_{\bot}$ 
for which $E_{\bot}=\hbar\omega_{\bot}$. Instead, 
choosing a space-time dependent $\eta$, one finds \cite{10} that 
$\eta$ satisfies the equation 
\beq 
\eta(z,t) = a_{\bot} \; 
\large( 1+2a_sN|f(z,t)|^2 \large)^{1/4} \; .  
\eeq 
By inserting this formula in the previous differential 
equation one gets a nonpolynomial Schr\"odinger equation (NPSE) 
\cite{10}, that reduces to the 1D GPE in the weakly-interacting 
limit $a_s N|f|^2 <<1$. 
The NPSE has been found to be very accurate in the descripition 
of BECs under transverse harmonic confinement 
and a generic longitudinal external potential. The NPSE is accurate 
with both repulsive and attractive scattering length \cite{8,10,11}. 
In our last paper \cite{12} we have extended the NPSE to include also 
beyond mean-field effects, like the formation of a gas of impenetrable 
bosons in the Tonks-Girardeau regime. 

\section{BEC bright solitons} 

For simplicity we consider first the case with $\omega_z=0$ 
in Eq. (3). In this case, it is well known that the 1D GPE 
(namely the Eq. (3) with $\eta = a_{\bot}$) 
with negative scattering length 
($a_s<0$) admits bright soliton solutions \cite{1}, that set up when 
the negative inter-atomic energy of the BEC compensates the positive 
kinetic energy such that the BEC is self-trapped. Scaling $z$ in units 
of $a_{\bot}$ and $t$ in units of $\omega_{\bot}^{-1}$, 
with the position \beq f(z,t)=\Phi(z-vt) e^{iv(z-vt)} 
e^{i(v^2/2 - \mu)t} \; , 
\eeq 
from 1D GPE one finds the text-book bright soliton  
\beq 
\Phi(z-vt)=\sqrt{\gamma\over 2} \; sech\left[{\gamma}(z-vt) \right] \; ,  
\eeq 
where $\gamma=|a_s|N/a_{\bot}$ and the chemical potential $\mu$ 
is given by $\mu= 1 - \gamma^2/2$, while the velocity $v$ of the bright 
soliton remains arbitrary. Given a value of the scattering length $a_s$, 
the number $N$ of condensed atoms fixes the chemical potential 
$\mu$ of the bright soliton. 
\par 
The solitary wave solution of Eq. (6), 
that we call {\it 1D bright soliton}, exists for any positive 
value of $\gamma$. On the other hand, it has been theoretically 
predicted \cite{10,13} and experimentally shown \cite{3,4} that above a 
critical interaction strength the BEC bright soliton, that we call 
{\it 3D bright soliton}, will collapse. 
To study the properties of a 3D bright soliton it is not necessary to solve 
the full 3D GPE; in fact, the 3D bright solitons can be very accurately 
described by NPSE \cite{8,10}. In particular, from the NPSE 
(namely the Eq. (3) with $\eta$ given by Eq. (4)) one finds the 
3D bright soliton written in implicit form 
$$ 
z - v t = {1\over \sqrt{2}} {1\over \sqrt{1-\mu}} \; 
arcth\left[ \sqrt{ \sqrt{1-2\gamma\Phi^2}-\mu \over 1-\mu } 
\right] 
$$
\beq-{1\over \sqrt{2}} {1\over \sqrt{1+\mu}} \; 
arctg\left[ \sqrt{ \sqrt{1-2\gamma\Phi^2}-\mu \over 1+\mu } \right] \; .   
\eeq 
It is important to observe that this equation is well defined only 
for $\gamma \Phi^2 <1/2$; at $\Phi^2 = 1/(2\gamma)$ the transverse 
width $\eta$ becomes zero. Moreover, by imposing the normalization 
condition, one finds \beq (1-\mu)^{3/2} - {3\over2} (1-\mu)^{1/2} 
+ {3\over 2 \sqrt{2}} \gamma = 0 \; ,  
\eeq ù
which admits solutions only for $0 < \gamma < 2/3$. 
\par 
There are thus two kind of collapse: the {\it transverse collapse}, 
when the condensate probability density $\rho=|\Phi|^2$ exceeds 
$\rho_c = 1/(2\gamma)$, and the {\it 3D collapse}, when $\gamma$ 
reaches the value $2/3$. Note that for a single 3D bright soliton 
at $\gamma_c=2/3$ the transverse size of the condensate soliton is 
not yet zero, in fact $\gamma_c \rho <1/2$. Nevertheless, if we consider 
two colliding 3D bright solitons with $\gamma<2/3$, at the impact the 
total density of the cloud can become equal to $\rho_c$ and the system 
will collapse. 

\begin{figure}
\centerline{\psfig{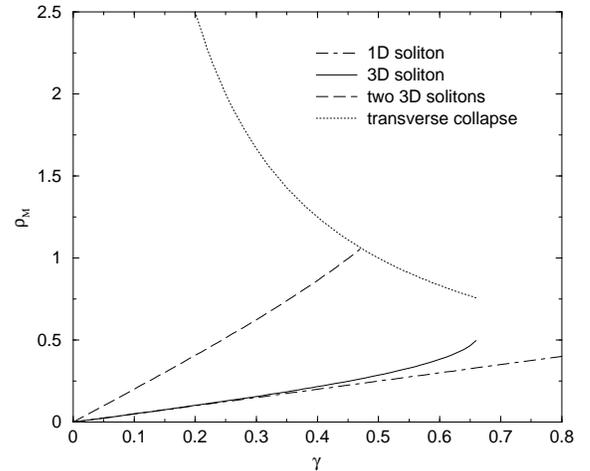}}
\caption{Maximum probability density $\rho_M$ of bright solitons as 
a function of the nonlinear strength $\gamma =|a_s|N/a_{\bot}$. 
Length $z$ in units $a_{z}=(\hbar /m\omega_z)^{1/2}$, density 
$\rho$ in units $1/a_{\bot}$, where $a_{\bot}=(\hbar /m\omega_{\bot})^{1/2}$ 
is the harmonic length of transverse confinement.} 
\end{figure} 

In Figure 1 we show the maximum probability density $\rho_M$ of bright 
solitons as a function of the strength $\gamma$. The solid line 
is the density of a single 3D bright soliton that ends at $\gamma =2/3$. 
The dot-dashed line is the probability density of the 1D bright soliton, 
that coincides with the solid line for small values of $\gamma$ 
(in fact, for $\gamma\ll 1$ Eq. (7) reduces to Eq. (6)). 
The dashed line is the maximum probability density of two colliding 
equal 3D bright solitons, 
that ends when the curve meets the dotted line, which is the density 
$\rho_c=1/(2\gamma)$ of the transverse collapse. Figure 1 shows that 
two colliding and equal 3D bright solitons produce a transverse collapse 
of the condensate for $\gamma =0.472$, which corresponds to a single 
soliton probability density equal to $0.265$. Only for lower values 
of $\gamma$ the collapse is avoided during collision and interference. 

\section{Bright soliton in the expulsive potential} 

We now consider the case $\omega_z\neq 0$ in Eq. (3) and 
try to simulate the experiment of the Ecole Normale 
Sup\'erieure \cite{4} with $^7$Li alkali-metal vapors. 
In that experiment a BEC with positive 
scattering length ($a_s>0$) has been condensed into a cigar-shaped 
harmonic trap. The scattering length has been then tuned to a small 
and negative value ($a_s<0$) via a Feshbach resonance. 
The resulting condensate, projected onto the expulsive 
harmonic potential of Eq. (1), has been observed to propagate 
over a distance of $1$ mm. 

\begin{figure}
\centerline{\psfig{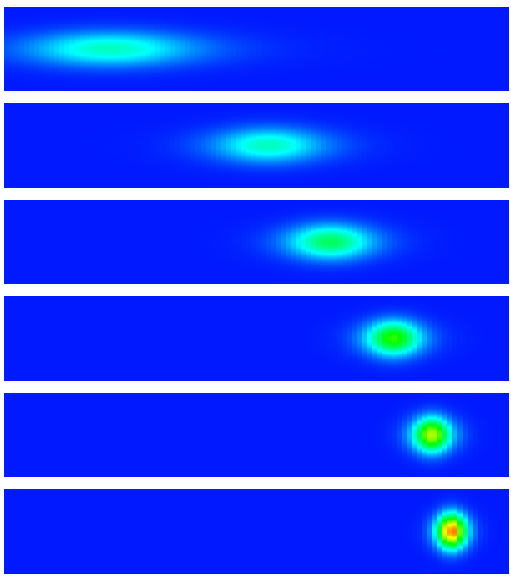}}
\caption{Density of the $^7$Li BEC in the expulsive 
potential obtained by solving the NPSE. The BEC cloud 
propagates over $1$ mm. 
Case with $a_s=0$ (ideal gas). There are 
$N=4\times 10^3$ atoms. Six frames from bottom to top: 
$t=2$ ms, $t=3$ ms, $t=4$ ms, $t=5$ ms, $t=6$ ms, 
$t=7$ ms. External harmonic potential given by 
Eq. (1). Red color corresponds to highest density.} 
\end{figure} 

\par 
Following the paper of Khaykovich {\it et al.} \cite{4}, 
we choose as initial condition for a fixed $a_s$ the ground-state 
of the stationary Eq. (3) with a fully confining 
harmonic potential: $\omega_{\bot} = 2\pi \times 710$ Hz and 
$\omega_z = 2\pi \times 50$ Hz. Then the time-dependent wave 
function is obtained by the numerical integration of the Eq. (3) 
with the expulsive potential: $\omega_{\bot} = 2\pi \times 710$ Hz but  
$\omega_z = 2\pi i \times 78$ Hz. 
Note that, as in the experiment, the initial position of 
the BEC cloud is shifted of $50$ $\mu$m on the left with 
respect to the maximum of the expulsive potential. 
A critical point is the choice of the number $N$ of atoms. 
To avoid the collapse 
we choose $N=4\times 10^3$, a value compatible 
with the experimental data and suggested by the 
variational theory \cite{13}. The NPSE is solved by using 
the finite-difference numerical algorithm described 
in \cite{14}. 

\begin{figure}
\centerline{\psfig{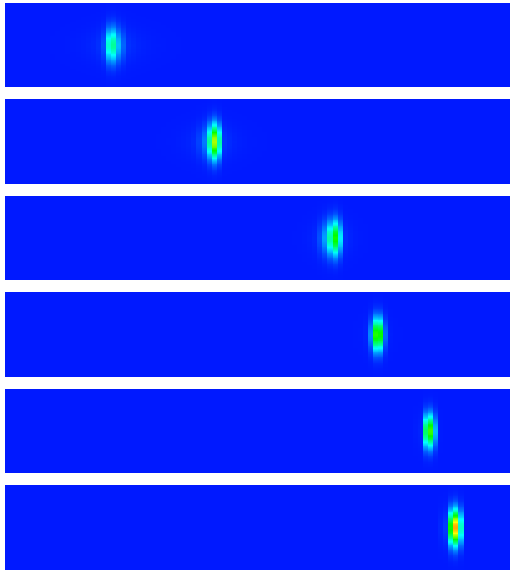}}
\caption{Density of the $^7$Li BEC in the expulsive 
potential obtained by solving the NPSE. 
The BEC cloud propagates over $1$ mm. 
Case with $a_s=-0.21$ nm (``bright soliton''). 
There are $N=4\times 10^3$ atoms. Six frames from bottom to top: 
$t=2$ ms, $t=3$ ms, $t=4$ ms, $t=5$ ms, $t=6$ ms, 
$t=7$ ms. External harmonic potential given by 
Eq. (1). Red color corresponds to highest density.} 
\end{figure} 

In Fig. 2 we plot six frames of the density of the BEC 
with $a_s=0$. This is the case of an ideal gas and the figure 
shows that the center of mass $z_{cm}$ of the bosonic cloud follows 
the law $z_{cm}(t)=z_0 \exp{(|\omega_z|t)}$. In addition, 
the longitudinal width of the cloud grows due to the 
dispersive kinetic term. In Fig. 3 we plot instead 
the time-dependent density of the BEC with $a_s=-0.21$ nm. 
Also here the center of mass $z_{cm}$ follows the exponential 
grow in time but the longitudinal width does not show 
an appreciable enlargement: the matter wave travels 
without spreading. It is important to stress that 
Fig. 2 and Fig. 3 are in remarkable good agreement with the 
absorption images shown in Fig. 3 of the experimental 
paper \cite{4} with the same values of the parameters. 
\par 
In their experiment, Khaykovich {\it et al.} \cite{4} have 
measured the root mean square size $\sigma$ of the longitudinal 
width versus the propagation time for three values of $a_s$: 
$a_s=0$, $a_s=-0.11$ nm, and $a_s=-0.21$ nm. 
Figure 4 shows their experimental data and our numerical results 
obtained with the NPSE. 
For $a_s=0$ the attractive interaction between atoms is zero 
and the expansion of the cloud is governed by the kinetic 
energy and the negative curvature of the longitudinal axial 
potential. For $a_s=-0.11$ nm the size $\sigma$ 
of the axial width is consistently below that of a non-interacting gas 
but the attractive interaction is not strong enough to 
avoid an appreciable enlargement of the cloud. 

\begin{figure}
\centerline{\psfig{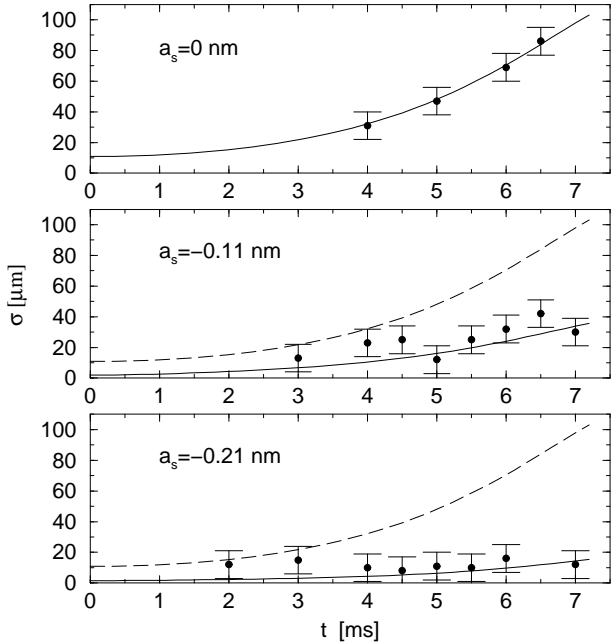}}
\caption{Root mean square size $\sigma$ 
of the longitudinal width of the BEC as a function 
of the propagation time $t$. The filled circles are 
the experimental data taken from Ref. [4]. The dashed 
line is the ideal gas ($a_s=0$) curve. 
The solid line is obtained from the numerical solution 
of the NPSE. } 
\end{figure} 

For $a_s=-0.21$ nm the enlargement is further reduced but our 
numerical results suggest that $\sigma$ is not truly constant. 
This is not surprising because a truly shape-invariant solitary 
wave is expected only in the absence of a longitudinal potential 
and with an appropriate initial condition. 
Nevertheless Fig. 4 shows that our results are fully compatible 
with the experimental data, which have the resolution limit 
$\Delta \sigma = 9$ $\mu$m of the imaging system. 
\par 
As previously discussed, in the calculations we have choosen 
$N=4\times 10^3$ to avoid the collapse of the BEC with 
$a_s =-0.21$ nm. In the experiments the collapse actually implies 
that a fraction of atoms is expelled from the Bose condensate via 
three-body recombination. This effect can be phenomenologically 
modelled by including a dissipative term in the GPE or in the NPSE, 
but a more rigorous treatment requires the solution of coupled 
time-dependent equations for the Bose condensate and 
the thermal cloud. 
\par 
Finally, we note that a sudden shift of the inter-atomic 
strength $|a_s|N/a_{\bot}$ to a very large value (for instance 
via a Feshbach resonance) produces a set of bright solitons 
(soliton train), which can travel in the expulsive potential. 
A detailed analysis of the formation of these multi-soliton 
configurations can be found in our recent papers \cite{8,15}. 

\section{Conclusions} 

We have shown that the 1D non-polynomial Schr\"odinger 
equation (NPSE) is able to accurately describe the 
time evolution of an attractive Bose-condensed cloud of 
$^7$Li atoms in an expulsive harmonic potential. 
Contrary to the 1D Gross-Pitaevskii equations, 
the NPSE takes also into account the space-time variations of the 
transverse width of the Bose-Einstein condensate and its numerical 
integration is much faster than the numerical solution of 
the 3D Gross-Pitaevskii equation. 
The theoretical results of the NPSE have been compared with 
the experimental data of a recent experiment performed at the 
Ecole Normale Sup\'erieure of Paris. Both experiment and theory 
show that, when the inter-atomic strength is sufficiently 
strong, a localized matter wave is obtained, which travels 
for a large distance without an appreciable spreading. 
Our numerical calculations suggest that this solitary wave 
is not fully shape-invariant but the experimental resolution 
does not allow one to clarify this point. 
It is important to stress that the limitations in 
the current resolution of the experimental imaging system 
can be overcome by creating a bright soliton 
with a larger longitudinal width. This can be achieved by using 
a smaller attractive inter-atomic strength in addition 
with a weaker expulsive potential. 
\par 
In conclusion, we observe that to make clear 
the particle-like nature of a soliton 
it is essential to investigate its interaction with another 
soliton. The interference of BEC bright solitons is an 
important subject which has to be analyzed in detail, 
having also implications for atom interferometry \cite{16}. 

\section*{Acknowledgments} 

We thank Alberto Parola, Davide Pini and Luciano Reatto for many 
useful discussions.

\end{document}